\documentclass[sn-mathphys-num]{sn-jnl}

\usepackage{graphicx}%
\usepackage{multirow}%
\usepackage{amsmath,amssymb,amsfonts}%
\usepackage{amsthm}%
\usepackage{mathrsfs}%
\usepackage[title]{appendix}%
\usepackage{xcolor}%
\usepackage{textcomp}%
\usepackage{manyfoot}%
\usepackage{booktabs}%
\usepackage{algorithm}%
\usepackage{algorithmicx}%
\usepackage{algpseudocode}%
\usepackage{listings}%

\theoremstyle{thmstyleone}

\theoremstyle{thmstyletwo}%

\theoremstyle{thmstylethree}%

\begin{document}

\title[Article Title]{3D Wasserstein generative adversarial network with dense U-Net based discriminator for preclinical fMRI denoising}

\author*[1]{\fnm{Sima} \sur{Soltanpour}}\email{simasoltanpour@cunet.carleton.ca}

\author[2]{\fnm{Arnold} \sur{Chang}}\email{chang.arn@northeastern.edu}
\equalcont{These authors contributed equally to this work.}

\author[3,4]{\fnm{Dan} \sur{Madularu}}\email{danmadularu@cunet.carleton.ca}
\equalcont{These authors contributed equally to this work.}

\author[2]{\fnm{Praveen} \sur{Kulkarni}}\email{p.kulkarni@northeastern.edu}
\equalcont{These authors contributed equally to this work.}

\author[2]{\fnm{Craig} \sur{Ferris}}\email{c.ferris@northeastern.edu}
\equalcont{These authors contributed equally to this work.}

\author[1]{\fnm{Chris} \sur{Joslin}}\email{chrisjoslin@cunet.carleton.ca}
\equalcont{These authors contributed equally to this work.}

\affil*[1]{\orgdiv{School of Information Technology}, \orgname{Carleton University}, \orgaddress{\street{1125 Colonel By Dr}, \city{Ottawa}, \postcode{K1S 5B6}, \state{Ontario}, \country{Canada}}}

\affil[2]{\orgdiv{Center for Translational NeuroImating (CTNI)}, \orgname{Northeastern University}, \orgaddress{\street{360 Huntington Ave}, \city{Boston}, \postcode{02115}, \state{Massachusetts}, \country{USA}}}

\affil[3]{\orgdiv{Department of Psychology}, \orgname{Carleton University}, \orgaddress{\street{1125 Colonel By Dr}, \city{Ottawa}, \postcode{K1S 5B6}, \state{Ontario}, \country{Canada}}}

\affil[4]{\orgdiv{Tessellis Ltd.}, \orgaddress{\street{350 Legget Drive}, \city{Ottawa}, \postcode{K2K 0G7}, \state{Ontario}, \country{Canada}}}

\abstract{Functional magnetic resonance imaging (fMRI) is extensively used in clinical and preclinical settings to study brain function, however, fMRI data is inherently noisy due to physiological processes, hardware, and external noise. Denoising is one of the main preprocessing steps in any fMRI analysis pipeline. This process is challenging in preclinical data in comparison to clinical data due to variations in brain geometry, image resolution, and low signal-to-noise ratios. In this paper, we propose a structure-preserved algorithm based on a 3D Wasserstein generative adversarial network with a 3D dense U-net based discriminator called, 3D U-WGAN.
We apply a 4D data configuration to effectively denoise temporal and spatial information in analyzing preclinical fMRI data. GAN-based denoising methods often utilize a discriminator to identify significant differences between denoised and noise-free images, focusing on global or local features. To refine the fMRI denoising model, our method employs a 3D dense U-Net discriminator to learn both global and local distinctions. To tackle potential over-smoothing, we introduce an adversarial loss and enhance perceptual similarity by measuring feature space distances.
Experiments illustrate that 3D U-WGAN significantly improves image quality in resting-state and task preclinical fMRI data, enhancing signal-to-noise ratio without introducing excessive structural changes in existing methods.
The proposed method outperforms state-of-the-art methods when applied to simulated and real data in a fMRI analysis pipeline.}


\maketitle

\section{Introduction}\label{sec1}

Preclinical fMRI provides valuable opportunities to connect non-invasive human fMRI to their biological origins \cite{zerbi2022use}. fMRI is a commonly utilized method for examining brain function by analyzing changes in blood oxygenation, blood flow, and blood volume signals. All these physiological indicators are coupled with neuronal activity, making it a fundamental component in fMRI studies of brain function. fMRI is a common technique that enables the identification of brain regions stimulated by physical activities or cognitive tasks. By indirectly measuring neural activity, fMRI can determine the functional activity in the brain by tracking changes in blood-oxygen levels \cite{yang2020robust}.
When using fMRI to study brain activity, researchers can identify active regions of the brain by determining the number of active voxels in each region or change in BOLD signal over time. The fMRI data inherently contains a large amount of noise due to physiological processes, technical limitations, and external interferences during image acquisition.
Denoising is one of the most significant steps in pre-processing of fMRI data. 
In clinical fMRI research related to the human brain, different denoising algorithms have been proposed and widely applied.  
These algorithms are divided into two different categories including conventional and learning-based methods.   

Methods based on independent component analysis (ICA) and principal component analysis (PCA) have been proposed as conventional methods for fMRI denoising.
ICA-AROMA \cite{pruim2015ica} is an automatic fMRI motion artifacts removal method based on ICA which first classifies independent components as noise or signal components and then regresses out the noise from fMRI data.
This method works based on regressing out certain components by manually or automatically labelling them as noise. 
Adding more regressors to fMRI data might help reduce noise, but it could also unintentionally remove the BOLD signal, leading to the loss of important information. 
Another challenge of ICA is that it necessitates a predefined number of components, which may be difficult to determine and can differ between various datasets and subjects.
Recently, Zhu et al. \cite{zhu2022denoise} proposed a multi-module PCA denoising algorithm based on random matrix theory for high resolution cat task fMRI data. 
A sparse decomposition method based on Morphological
Component Analysis (MCA) is proposed by Nguyen et al. \cite{nguyen2022morphological}.  This method leverages sparse representations to detect functional connectivity for clinical resting-state (rsfMRI) and task fMRI data. 
Although, these approaches provide a valuable contribution to the field of fMRI denoising, their application for denoising preclinical rat fMRI data is limited due to differences in brain size, spatial and temporal resolution between human and preclinical data.
 
Recently, artificial intelligence and deep learning has attracted researches attention in neuroscience research.
A deep neural network (DNN) is proposed by Yang et al. \cite{yang2020robust} to remove noise in task fMRI data without the need for explicitly modelling the noise.
DNN is also used by Theodoropoulos et al. \cite{theodoropoulos2021automatic} to present advanced DNN architecture to classify noise by applying both spatial and temporal information in rsfMRI data. 
A deep attentive spatio-temporal feature learning is proposed by Heo et al. \cite{heo2022deep} for noise-related component identification in rsfMRI. This algorithm decomposes data into multiple components and then noise-related components are regressed out. A unified deep attentive spatio-spectral temporal feature fusion framework has been proposed by Lim et al. \cite{lim2024unified} for rsfMRI denoising. However, these techniques have been developed for human brain data and present challenges for preclinical applications due to differences in image resolution, and contrast. Moreover, these methods are limited to processing task fMRI, or rsfMRI data and have not been applied across different datasets. In this way, a preclinical denoising approach which can prevent over smoothing, and blurring of anatomical details in both preclinical rsfMRI and task fMRI is essential.

Nowadays, with deep learning advancements, Generative Adversarial Networks (GANs) \cite{goodfellow2014generative} have been applied for denoising different types of medical imaging such as RED-WGAN for denoising 3D MRI images \cite{ran2019denoising},
3D GAN for dose reduction in 4D cerebral CT Perfusion (CTP) imaging \cite{moghari2021efficient}, 
DU-GAN for low-dose  computed tomography (LDCT) denoising \cite{huang2021gan}, RIRGAN for brain MRI denoising \cite{yu2023rirgan}, DISGAN for MRI super-resolution and cleaning \cite{wang2023disgan}, MDGAN to reduce speckle in optical coherence tomography images \cite{yu2023multiscale}, and content-noise complementary GAN for LDCT denoising with local and global discriminator \cite{sarkar2024noise}. 
GANs \cite{goodfellow2014generative} and their variants have been also applied to various preclinical MRI tasks such as preclinical MRI motion correction \cite{bao2022retrospective}, and MouseGAN \cite{yu2021mousegan} for mouse brain segmentation. 
However, to the best of our knowledge this is the first time that GAN is applied for preclinical fMRI denoising, encompassing the joint capture of both temporal and spatial information. 
 
In this paper, we propose a novel preclinical fMRI denoising algorithm which applies a 3D GAN model in the context of whole-brain 4D fMRI data. We introduce a dense 3D U-Net based discriminator within the WGAN framework called 3D U-WGAN.
Our approach focuses on noise reduction, aiming to predict high-quality denoised fMRI data from noisy acquisitions. The U-net-inspired discriminator in our model significantly enhances the efficiency of estimating denoised fMRI data across the entire brain when compared to other methods.
The proposed GAN-based fMRI denoising algorithm is designed to effectively reduce noise across preclinical rsfMRI and task fMRI datasets. Our method does not depend on domain-specific assumptions or handcrafted features by exploiting the power of generative adversarial networks in order to adapt to different noise patterns and data characteristics. The algorithm learns to recognize normal neural signals from different degrees of noise through adversarial training. 

The remainder of this paper is organized as follows: Section~\ref{sec 2} describes the proposed preclinical fMRI denoising framework, 3D U-WGAN. The databases used in this study are detailed in Section~\ref{sec 3}. Section~\ref{sec 4} presents the experimental results and evaluations. Finally, conclusions are provided in Section~\ref{sec 5}. 

\section{Proposed Method}

\label{sec 2}
\begin{figure}
\begin{center}
\begin{tabular}{c}
\includegraphics[height=8.5cm]{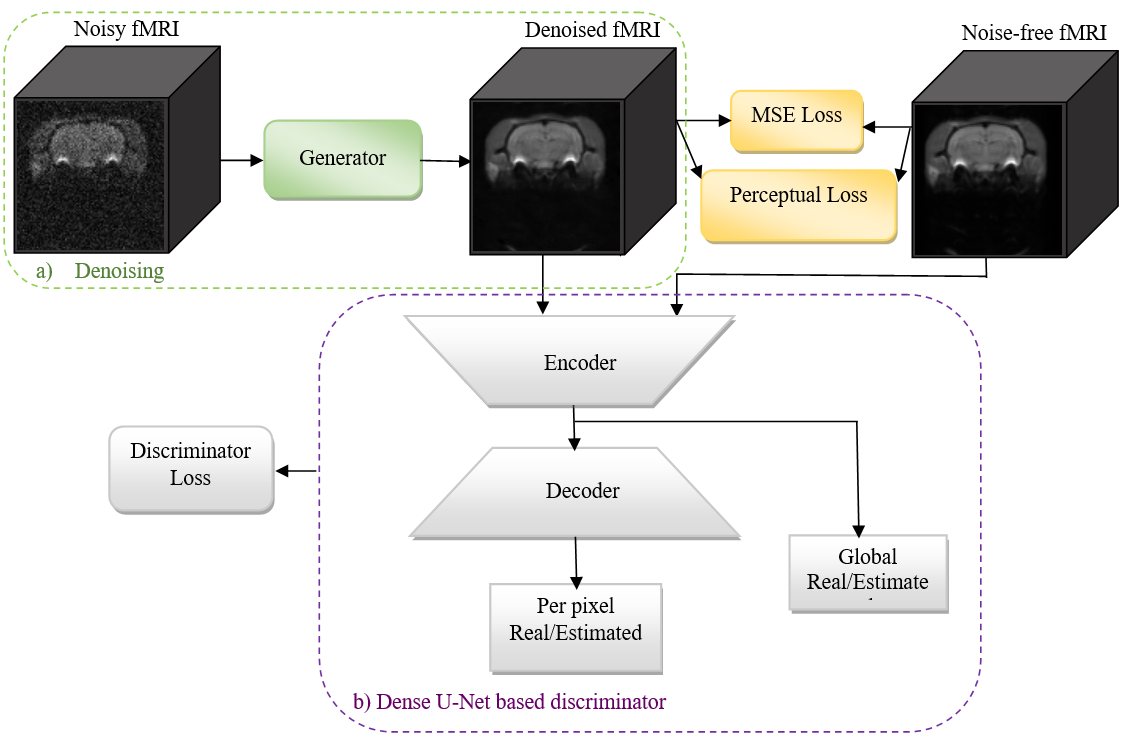}
\end{tabular}
\end{center}
\caption 
{ \label{Fig. 1}
Overall framework of 3D U-WGAN for fMRI denoising. (a) Denoising process: the generator generates denoised fMRI. (b) The dense U-Net based discriminator provides both general structure and local feedback to the generator. } 
\end{figure}

The overall framework of the proposed 3D U-WGAN for fMRI denoising is presented in Fig.~\ref{Fig. 1}. The model contains a generator and a dense U-Net based discriminator.
The generator network receives a noisy fMRI data and produces an estimated fMRI data that closely resembles the corresponding real noise-free fMRI data. The discriminator network receives a set of images, including both the estimated denoised fMRI and the corresponding real noise-free fMRI data. Its objective is to distinguish between the real and estimated pairs. 
If the discriminator is able to easily differentiate between the estimated fMRI and the real noise-free image, it indicates that the estimated image lacks resemblance to the real one. In this case, the generator needs to improve its performance and generate more realistic estimations. On the other hand, if the discriminator struggles to distinguish between them, it suggests that the discriminator should be enhanced to better differentiate the images. Therefore, the two networks have opposing tasks: the generator aims to synthesize a noise-free fMRI, while the discriminator aims to discern the estimated image from the real noise-free image. To train the generator network, we take into account not only the feedback from the discriminator but also the estimation error loss. 
Literature shows that loss function in deep learning-based denoising methods is more important than the network architecture as it affects on the image quality \cite{zhao2016loss}.
To efficiently handle the loss function, in this work, inspired by Ran et al. \cite{huang2021gan}, an adversarial loss related to the discriminator is applied in combination with mean-squared error (MSE) loss and perceptual loss.

To enhance the regularization of the denoising model during adversarial training we apply a U-Net \cite{ronneberger2015u} based discriminator along with the WGAN \cite{arjovsky2017wasserstein} framework.
By employing the U-WGAN, the denoising model can effectively learn both the overall and local distinctions between the denoised and noise-free images.
Specifically, our proposed model utilizes a dense U-Net architecture \cite{kolavrik2019optimized} for the discriminator, which consists of both an encoder and a decoder network. The encoder plays a role in encoding the input into a singular scalar value, highlighting the overall structures of the images. Simultaneously, the decoder reconstructs a confidence map on a pixel-level basis, capturing the local variations between the denoised image and the noise-free image.

\subsection{Noise Reduction Process}
The objective of fMRI denoising is to produce a high-quality fMR image from a noisy fMR image. $I_N \in R^{m\times n \times q\times t}$ represents the noisy fMR image, and $I_{NF} \in R^{m\times n\times q\times t}$ represents the corresponding noise-free fMR image. The size of each slice is defined by $m\times n$, $q$ is the number of slices, and $t$ is the number of time points. Their relationship can be expressed as: $I_N = \delta(I_{NF})$, where $\delta$ is a mapping function that represents the noise contamination. Since deep learning-based methods are independent of the statistical properties of the noise and function as black boxes, fMRI denoising can be simplified to finding the optimal approximation of the inverse function $\delta^{-1}$. 
The denoising process is formulated in Eq.~\ref{eq. 1}.

\begin{equation}
\label{eq. 1}
argmin_f\Vert \hat{I_{NF}}-I_{NF}\Vert_2^2
\end{equation} 
 
where $\hat{I_{NF}}=f(I_N)$ is the estimation of $I_{NF}$, and $f$ represents the optimal approximation of $\delta^{-1}$.

\subsection{WGAN}
In this work, we adopt the improved version of WGAN with a gradient penalty \cite{gulrajani2017improved} to accelerate the convergence instead of conventional GANs. This choice helps stabilize the training process and enhances the visual quality of denoised fMRI images.
The training procedure of WGAN can be described as a minimax game \cite{gulrajani2017improved} by solving the following equation:

\begin{align}
\label{eq. 2}
\min_G \max_D L_{WGAN}(D, G) = -E_{I_{NF} \sim P_{NF}} [D(I_{NF})] +E_{I_N \sim P_N} [D(G(I_N))]\nonumber \\
+\lambda \mathbb{E}_{\hat{I_{NF} } \sim P_{\hat{I_{NF} }}} \left[ \left( \|\nabla_{\hat{I_{NF} }} D(\hat{I_{NF} })\|_2 - 1 \right)^2 \right]
\end{align}

where the first two terms estimate the Wasserstein distance, the last term serves as the gradient penalty for regularizing the network. Here, $\hat{I_{NF}}$ is uniformly sampled along straight lines between pairs of generated and real samples, and $\lambda$ is a constant that controls the weighting of the penalty.

In statistical terms, $I_N$ and $I_{NF}$ can be seen as two sets of data taken from separate distributions: the noisy image distribution $P_N$ and the noise-free image distribution $P_{NF}$, respectively. The denoising process involves a mapping procedure where one distribution is transformed into another. This means that the function $f$ converts the samples from $P_N$ to a different distribution $P_G$ (the generated image distribution) that closely resembles $P_{NF}$. Therefore, the objective of GANs is to train a generator distribution $P_G$. To achieve this, GANs utilize a generator network G that takes the noisy input from a probability distribution $P_N$ and transforms it into a generated sample.
The generator is trained by competing against a discriminator network D, which aims to distinguish between samples from the actual data distribution $P_{NF}$ and the generated distribution $P_G$. 

\subsection{3D Dense U-Net Based Discriminator}
Denoising based on GAN such as 3D MRI denoising by Ran et al. \cite{ran2019denoising}, and CTP denoising by Moghari et al. \cite{moghari2021efficient} typically maintains the competition between the generator and discriminator at the structural level. 
However, the discriminator faces a challenge in retaining previous samples due to the shifting distribution of synthetic samples caused by the continuous changes in the generator during training. As a result, it fails to maintain a robust data representation capable of capturing both global and local differences in images \cite{schonfeld2020u}. To address this challenge, we introduce a novel approach by employing a 3D dense U-Net based discriminator for fMRI denoising. Our method incorporates a 3D U-Net architecture, which consists of an encoder, a decoder, and several dense interconnections that facilitate faster learning and enhance the level of details in the denoising process to effectively address the task of fMRI denoising. 

U-Net has demonstrated effective results in many image translation \cite{schonfeld2020u} and segmentation \cite{azad2024medical} tasks . In the context of medical image denoising, 3D U-Net has been applied in generator architecture for PET denoising \cite{wang20183d} and CTP denoising \cite{moghari2021efficient}to synthesize the noise-free image. This deep architecture incorporates skip connections to combine hierarchical features, allowing for the integration of multi-level information and enhancing the generation of high-quality images. The U-Net based discriminator containing an encoder, a decoder, and several skip connections has been applied by Huang et al. \cite{huang2021gan} for 2D Low-dose computed tomography (LDCT) denoising in both image and gradient domains. 

In this paper we adopt a 3D dense U-Net to replace the standard discriminator classification in GANs which has not been explored as the discriminator for denoising. We utilize the U-Net architecture to substitute the conventional classification-based discriminator within GANs, creating a U-Net-style discriminator. This adaptation enables the discriminator to maintain both global and local data structure.
The encoder component of the discriminator network ($D$), denoted as $D_{enc}$, follows a conventional discriminator structure that progressively reduces the input dimensions using multiple convolutional layers. This allows the network to capture the global structural context of the input data. In contrast, the decoder component, $D_{dec}$, performs a progressive upsampling process. Both components integrate densely connected layers to process 3D data. This design choice further enhances the discriminator's ability to capture local details in both noise-free and noisy samples. Moreover, the discriminator loss is computed from the outputs of both $D_{enc}$ and $D_{dec}$, whereas previous works such as 3D MRI denoising by Wang et al. \cite{wang20183d}, and 3D MRI denoising by Rang et al. \cite{ran2019denoising} primarily used a traditional discriminator that only classified inputs as noise-free or noisy based on the encoder. 
The U-Net-based discriminator provides the generator with more comprehensive feedback, incorporating both local per-pixel details and global structural information.

\subsection{Loss Functions}
In this section, we introduce the generator loss function which is used to optimize the generator to generate denoised fMRI data. The final loss function consists of MSE loss, perceptual loss, and discriminator loss.

\subsubsection{MSE Loss} 
To encourage the generation of denoised fMRI data that closely resemble the noise-free images in terms of pixel-level characteristics, we employ a pixel-wise loss between the noise-free fMRI and denoised fMRI images to minimize pixel-wise differences between them. The MSE loss is calculated by the following formula:

\begin{equation}
L_{MSE} = E_{(I_{N},I_{NF})}\Vert {I_{G}}-I_{NF}\Vert_2^F
\label{eq. 3}
\end{equation} 
where $E$ and $\Vert \Vert_2^F$ represents the average value and Frobenius norm respectively.

\subsubsection{Perceptual Loss} 
Although the MSE loss function can yield a high peak signal-to-noise ratio (PSNR), it might result in a loss of fine details \cite{ledig2017photo}. To address this, we incorporate a perceptual loss function that operates within a feature space \cite{yang2018low}. By utilizing this approach, we can obtain a more precise representation of image features, leading to the preservation of crucial details in the output as shown in 
Eq. \ref{eq. 4}. 

\begin{equation}
L_{Per} = E_{(I_{N},I_{NF})}\Vert {\phi({I_{G}}})-\phi({I_{NF}})\Vert_2^F
\label{eq. 4}
\end{equation}

where $\phi$ is a feature extractor.
The feature extractor utilized in our approach is a pre-trained VGG network \cite{simonyan2014very}. The VGG-19 network comprises 16 convolutional layers followed by 3 fully-connected layers. The output obtained from the 16th convolutional layer serves as the feature extracted by the VGG network, which is then employed in the perceptual loss function \cite{yang2018low}. The input to the VGG network is 2D slices from 3D data to extract features and calculate the perceptual loss.

\subsubsection{Discriminator Loss} 
The calculation of the adversarial loss relies on the discriminator, which is a classification network designed to acquire knowledge about distinguishing between denoised images and noise-free images. The discriminator is capable of assessing the most distinctive variation either on a broader scale encompassing the entire image or on a smaller scale focusing on specific local regions. This means that a single unit of the discriminator's output can correspond to either the entire image or a specific local region, depending on the context. 
The discriminator loss is defined in the context of WGAN, Eq.\ref{eq. 2}.

The discriminator loss, $L_D$, considering inputs from both $D_{enc}$ and $D_{dec}$, is expressed in Eq. \ref{eq. 5}.
\begin{align}
\label{eq. 5}
L_D = -E_{I_{NF} \sim P_{NF}} [D_{enc}(I_{NF})+D_{dec}(I_{NF})] +E_{I_N \sim P_N} [D_{enc}(I_G)+D_{dec}(I_G)]\nonumber \\
+\lambda \mathbb{E}_{\hat{I_{NF} } \sim P_{\hat{I_{NF} }}} \left[ \left( \|\nabla_{\hat{I_{NF} }} (D_{enc}(\hat{I_{NF} })+D_{dec}(\hat{I_{NF} }))\|_2 - 1 \right)^2 \right]
\end{align}

where $I_G$ denotes the denoised image which is the generator output, $D_{enc}$ represents the global adversarial  and $D_{dec}$ represents the local adversarial.

\subsubsection{Final Loss} 
The final loss function consists of MSE loss, perceptual loss, and discriminator loss. It is applied to optimize the model and defined by the following equation.

\begin{equation}
L_{3D U-WGAN} = \lambda_{MSE}L_{MSE}+\lambda_{Per}L_{Per}+\lambda_{D}L_{D}
\label{5}
\end{equation}
%


where $\lambda_{MSE}$, $\lambda_{Per}$, and $\lambda_{D}$ are the weights for $L_{MSE}$, $L_{Per}$, and $L_{D}$ respectively. 

\subsection{Network Architecture}
The proposed method adopts the GANs framework to efficiently optimize the generator for fMRI denoising with the dense U-Net  discriminator. The discriminator is based on the U-Net architecture, which enables it to capture both global structures and local details. In this section, we provide a detailed description of the network architectures for both the generator and the U-Net-based discriminator.
\subsubsection{Generator Architecture}
\begin{figure}
\begin{center}
\begin{tabular}{c}
\includegraphics[height=2.5 cm]{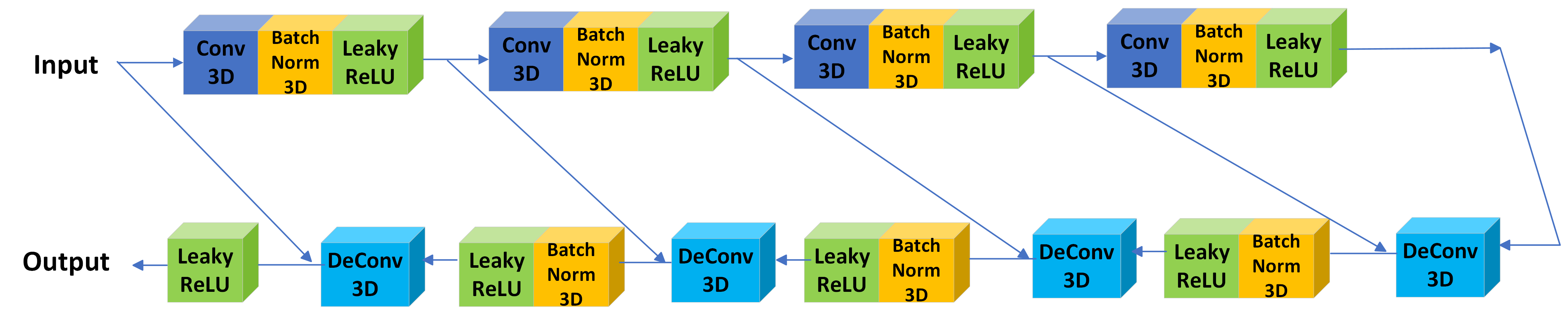}
\end{tabular}
\end{center}
\caption 
{ \label{Fig. 2}
Encoder-decoder based generator architecture.} 
\end{figure}
For fMRI denoising within our framework, we utilize the generator of RED-WGAN \cite{ran2019denoising} as shown in Fig. \ref{Fig. 2}.  
The generator G employs an encoder-decoder structure consisting of a total of 8 layers. These layers include 4 convolutional layers and 4 deconvolutional layers. Each convolutional layer is connected to its corresponding deconvolutional layer through short connections. With the exception of the final layer, all layers perform a sequence of operations: 3D convolution, batch normalization, and Leaky rectified linear unit (LeakyReLU) operation. The final layer, on the other hand, only performs a 3D convolution and a LeakyReLU operation. In this paper, all kernels are set to a size of 3x3x3, and the number of filters used follows the sequence: 32, 64, 128, 256, 128, 64, 32, 1.

\subsubsection{Discriminator Architecture}
\begin{figure}
\begin{center}
\begin{tabular}{c}
\includegraphics[height=10.5 cm]{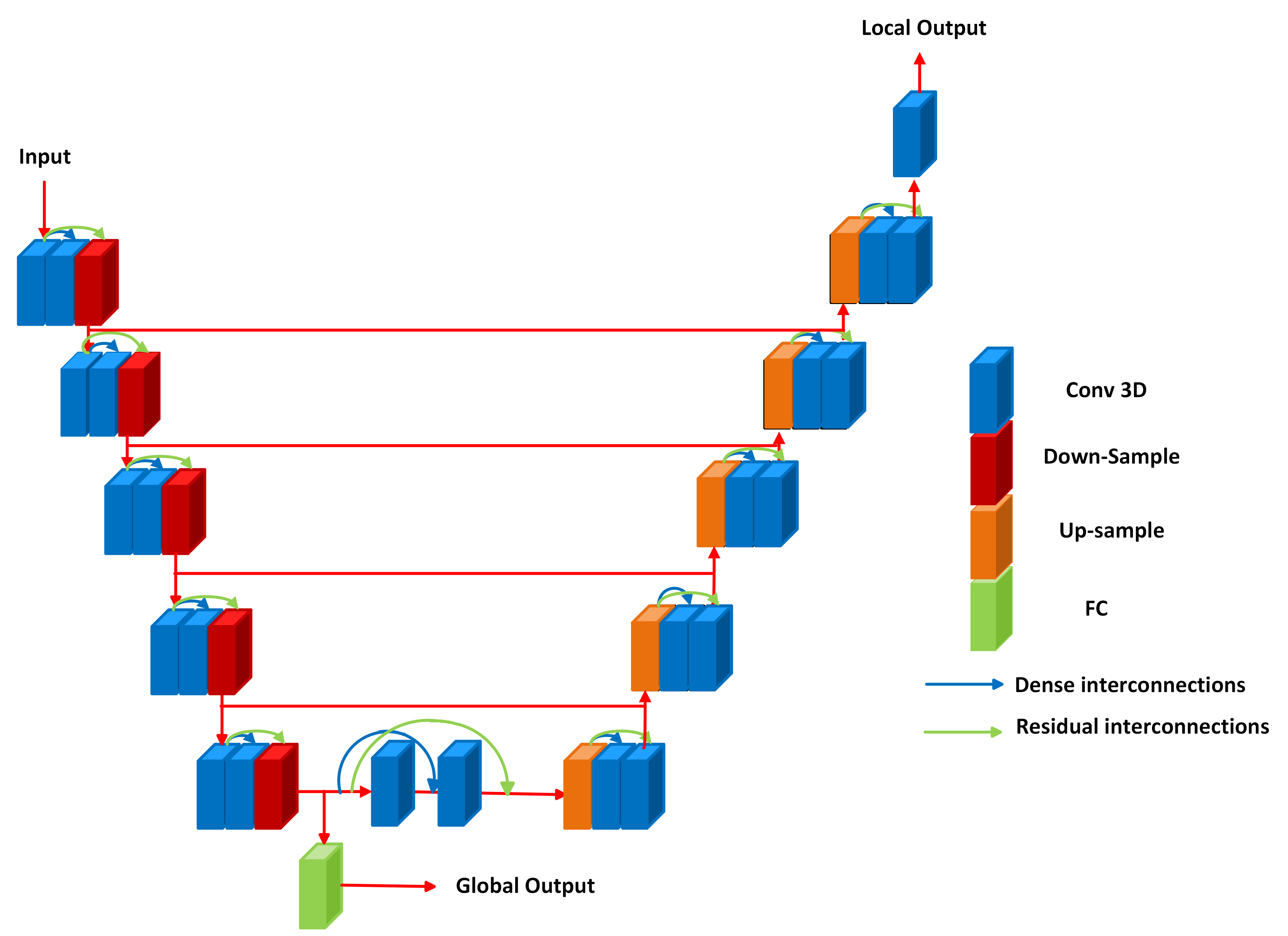}
\end{tabular}
\end{center}
\caption 
{ \label{Fig. 3}
3D dense U-Net based discriminator architecture.}
\end{figure}

In this section, we introduce a 3D dense U-Net based discriminator, which builds upon the original 3D U-Net version \cite{cciccek20163d}. The 3D dense U-Net \cite{kolavrik2019optimized} incorporates additional interconnections between layers that process the same feature size, as illustrated in Fig. \ref{Fig. 3}. The main goal is to explore the concept of dense interconnections in U-Net-type networks for denoising tasks. This approach draws inspiration from the principles commonly employed in deep neural networks for classification. 
The uniqueness of this research lies in its ability to process preclinical image data in its original resolution, surpassing the performance of standard 3D U-Net models in terms of accuracy. The incorporation of interconnections in the network aids in achieving a faster learning curve and capturing more intricate details in the images.

The discriminator architecture has been illustrated in Fig. \ref{Fig. 3} and shows the encoder module consists of five down-sampling layers that progressively increase the number of filters, starting with 32, followed by 64, 128, 256, and 512. At the lowest point of the encoder module, a fully connected layer is employed to generate the global confidence score. Similarly, the decoder module utilizes the same number of layers in reverse order. It processes the upsampled features. Following this, a convolutional layer is utilized to generate the per-pixel confidence score. In each up-sampling block, the feature size is increased by a factor of two at the start. This is achieved using a transposed convolution layer, with a stride size of 2 for each dimension. Each convolutional layer of encoder and decoder except the last one is followed by a spectral normalization layer \cite{miyato2018spectral} and a leaky ReLU activation function.

\subsection{Implementation Details}
To train the 3D U-WGAN, we used the high-performance computing (HPC) facilities within the Digital Research Alliance of Canada \cite{HPC}. An Adam optimization method \cite{kingma2014adam} is used to optimize the networks in the proposed algorithm
with a learning rate of $10^{-4}$. 
The weights of the model are updated over mini-batches of size 32 with batch normalization. We choose $\lambda=10$ for the discriminator loss as suggested in \cite{gulrajani2017improved}. The loss functions' hyperparameters including $\lambda_{MSE}$, $\lambda_{Per}$, and $\lambda_{D}$ were set experimentally to 1, 0.1, 0.2 respectively. These hyperparameters are established in a step-by-step manner. Initially, when using just pixelwise loss, this approach results in rapid convergence, focusing solely on minimizing the MSE loss can result in excessive smoothing and produce blurry outcomes, which in turn leads to the loss of fine structural details. We fix the value of the parameter $\lambda_{MSE}$ at 1. In the second step, we fine-tune the parameter $\lambda_{Per}$ to effectively capture feature information. Ultimately, we adjust the parameter $\lambda_{D}$ to capture intricate details through the adversarial loss, we initiate by assigning a low value to the parameter $\lambda_{D}$. Subsequently, we systematically elevate the significance of the adversarial loss and observe the outcomes in denoising.

\section{Center for Translational NeuroImaging (CTNI) preclinical fMRI Dataset}
\label{sec 3}
This dataset was collected by Center for Translational NeuroImaging (CTNI) of Northeastern University.
Experiments were conducted using a Bruker Biospec $7.0T/20-$cm USR horizontal magnet (Bruker, Billerica, Massachusetts) and a $20-G/$cm magnetic field gradient insert ($ID = 12$ cm) capable of a 120-$\mu$s rise time (Bruker). Radiofrequency signals were sent and received with the quad-coil electronics built into the animal restrainer.

At the beginning of each imaging session, a high-resolution anatomical data set was collected using the Rapid Acquisition with Relaxation Enhancement (RARE) pulse sequence (25 slice, 1 mm, Field of View (FOV) 3.0 cm, $256\times256$, TR 2.5 sec, TE 12.4 msec, NEX 6, 6-minute acquisition time). Functional images for the task fMRI portion will be acquired using a Half-Fourier acquisition, single-shot, turbo-spin echo sequence. A single scanning session was acquired $96\times96$ in-plane resolution, 22 slices every 6 seconds (TR = 6000 msec, TE-eff = 48 msec, RARE factor = 36, NEX 1) repeated 100 times for every 10-minute scan. rsfMRI data containing 108 subjects was collected before and after task fMRI scanning using a spin-echo triple-shot echo-planar imaging (EPI) sequence (imaging parameters: matrix size $96\times96\times22$ (length $\times$ width $\times$ depth), repetition time 1000 ms, (Eff-TR: 3000 ms) echo time 15 ms, voxel size $0.312\times0.312$ mm, slice thickness 1.2 mm, 300 repetitions, acquisition time 15 min).

\subsection{4D fMRI Data Configuration}
\begin{figure}
\begin{center}
\begin{tabular}{c}
\includegraphics[height=6.5 cm]{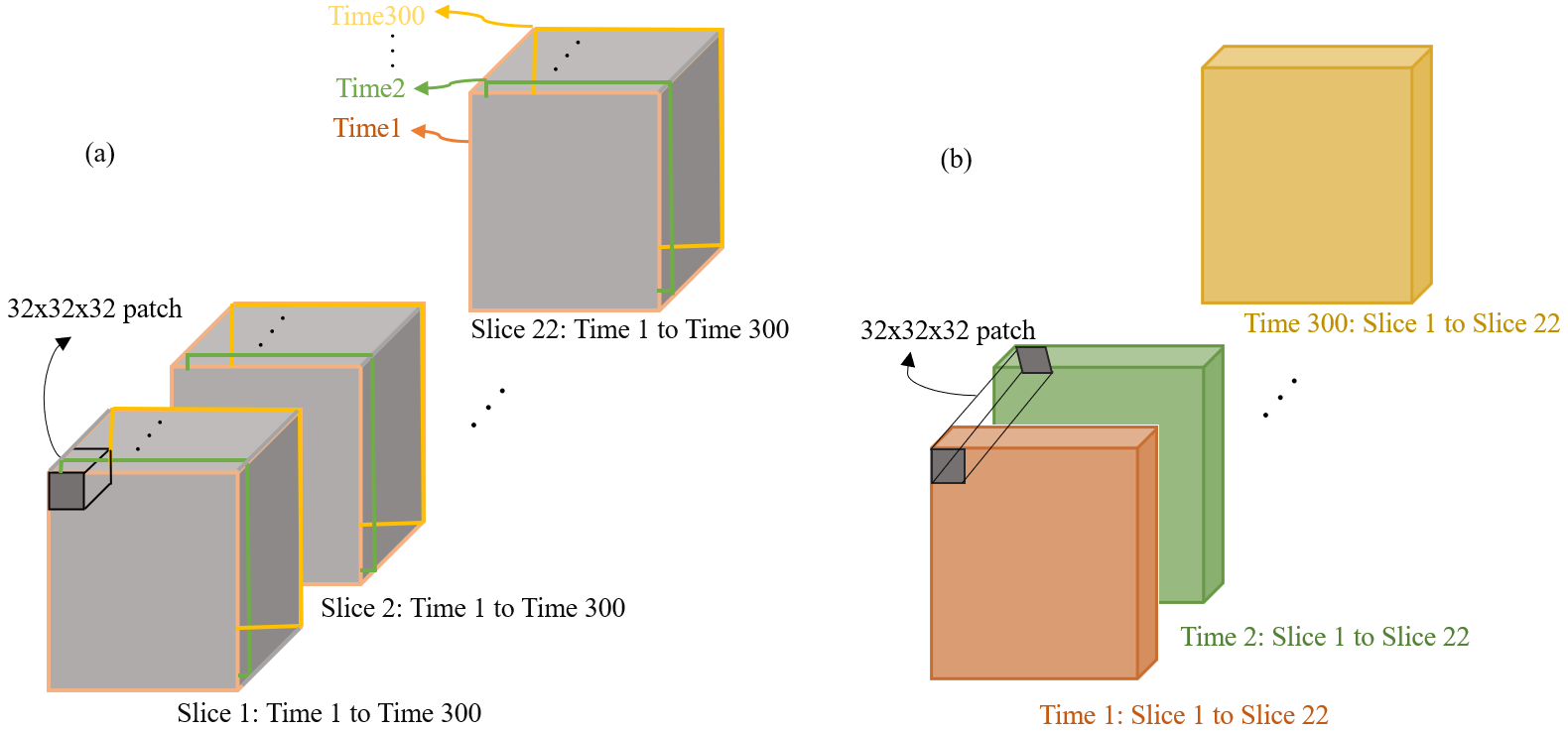}
\end{tabular}
\end{center}
\caption 
{ \label{Fig. 4}
3D input patches from 4D fMRI data by a) concatenating volumetric slices (slice-based), b) concatenating time sequences (time-based).} 
\end{figure}
Deep learning approaches typically require a substantial amount of training data, which can be challenging to obtain, particularly in preclinical settings. To overcome this issue, we extracted voxels patches from the samples to train the network. 
This approach has demonstrated effectiveness in improving the detection of perceptual differences while significantly increasing the number of available samples.
We applied and compared two approaches \cite{moghari2021efficient} for utilizing the four-dimensional characteristics of fMRI data in the context of the 3D GAN model as shown in Fig. \ref{Fig. 4}. 
The first approach illustrated in Fig. \ref{Fig. 4}(a) entailed combining the 300 volumetric frames from a sample and extracting 32 x 32 x 32 patches from the merged dataset, which had dimensions of 96 x 96 x 6600, using a stride of 32. This process resulted in 1863 image patches for each training/testing sample, distributed as 3 x 3 x 207 in the x, y, and z dimensions, respectively. It's worth noting that there was no overlap between these patches, and for the final 32 patches along the z-axis, zero-padding was applied. We term this data organization as slice-based, and we refer to the model trained with this configuration as 3D UWGAN$_{xyz}$.

The second approach depicted in Fig. \ref{Fig. 4}(b) entailed combining the time series data for each 2D brain slice. These time sequences, organized by brain slices, were subsequently concatenated to create a 3D dataset with dimensions of 96 x 96 x 6600. From this dataset, we extracted spatio-temporal image patches of size 32 x 32 x 32. We denote this data organization as time-based and the model trained with this data arrangement as 3D UWGAN$_{xyt}$.

Conceptually, the fMRI signal fluctuation in the plane reflects spatial information within a slice in both the slice-based and time-based data configurations. The fMRI signal fluctuation along the 3rd dimension of the slice-based configuration prioritizes the spatial information from adjacent slices within a frame; for the time-based configuration, it prioritizes the temporal information related to the dynamics of fMRI signal across scanning frames. The  network was trained and tested on the two data arrangements separately to compare the performance difference with respect to this trade-off.

\subsection{Data for training and test}
\label{3.2}
We employed a five-fold cross-validation approach to train and evaluate the network under varying experimental conditions, including different noise levels and 4D data arrangements. In each of the five cross-validation splits, the subjects were divided into five sets, with four sets (comprising 100 subjects) used for training and one set (with 8 subjects) for testing. Following the method outlined in the previous section, the total number of patches used for training and testing was 186,300 (1863 patches per subject $\times$ 100 subjects) and 14,904 (1863 patches per subject $\times$ 8 subjects) for both slice and time-based configurations respectively. All the estimated image patches corresponding to a particular configuration were combined to create the complete estimated fMRI volume.

Noisy fMRI data is simulated by adding Rician noise \cite{ran2019denoising} in different standard deviation ($\delta$) values of the noise. Rician noise is often used in the context of fMRI denoising because it is a statistical model that closely approximates the characteristics of noise present in fMRI data \cite{zhu2022denoise}.

\subsection{Data simulation for fMRI analysis}
\label{sec 3.3}
The preprocessing steps were conducted using a combination of software tools, including AFNI (v17.1.12), FSL (v5.0.9), DRAMMS (v1.4.1), and MATLAB. Brain tissue masks were manually delineated using 3DSlicer and applied to perform skull-stripping. Motion outliers and spikes were identified and regressed out, followed by slice timing correction. Head motion correction was applied using the first volume as a reference, and images were normalized
through affine registration to the 3D MRI Rat Brain Atlas. The atlas provided 173 annotated brain regions for segmentation. Quality assurance measures were carried out, along with band-pass filtering $(0.01 \, \text{Hz} \sim 0.1 \, \text{Hz})$ to reduce drift and noise,
detrending, spatial smoothing (full width at half $\text{maximum} = 0.8 \, \text{mm})$, and nuisance regression using motion outliers, motion parameters, and mean white matter and cerebrospinal fluid time series as regressors. 
The input data consisted of anesthesized rat fMRI data with no condition or stimulus applied. 

We then generated fMRI phantoms with spatial activations modeled by Gaussian functions and temporal modulations convolved with a rat-specific hemodynamic response function (HRF).  Activation centers were defined for three regions of interest (ROI) in the rat brain: the superior colliculus, the primary visual cortex (V1), and the secondary visual cortex (V2). Each center was described by its 3D coordinates and an initial amplitude, and together were chosen to represent a visual processing task. These positions and amplitudes were used to simulate Gaussian-shaped activations within the brain. The peak HRF amplitudes were set to 10\% of the baseline at the center of the activation pattern and gradually decreased with distance according to a Gaussian distribution. 
 
A rat-specific HRF \cite{lambers2020cortical} was employed to model the temporal modulation of the activations. Each activation followed an on-off cycle of 30 seconds on and 60 seconds off, repeated for 10 cycles. The modulation pattern was convolved with the HRF to create smooth transitions between the on and off periods. For each time point, activations were generated as Gaussian functions centered at the predefined locations. These Gaussian activations were scaled by the HRF-modulated amplitude corresponding to each time point. A standard deviation of 4 voxels was used to represent the spatial spread of the activations. Noise was added to the activation data based on a signal-to-noise ratio (SNR) of 30 dB. Gaussian-distributed noise was scaled appropriately to achieve the desired SNR, ensuring that the simulated data closely resembled real fMRI data in terms of noise characteristics. Additionally, a binary mask corresponding to the activation centers, each with a radius of 1 standard deviation (4 voxels), was created to serve as the ground truth.

After the fMRI phantom generation, a general linear model (GLM) analysis was performed using SPM (Statistical Parametric Mapping) to identify the brain regions associated with the simulated activations. The following procedure was followed for GLM analysis: The model specification included a repetition time (TR) of 3 seconds and an HRF model to represent the expected temporal response. For each image, session-specific information was defined, including the scan data and the onset and duration of the stimulus (activation) periods. Model estimation was performed for each phantom using the classical (ordinary least squares) method. A contrast representing the effect of the stimulus (activation) was defined, with a weight of 1 applied to the stimulus regressor, allowing us to estimate the statistical significance of the activations. After running the GLM analysis, the output t-statistic image was generated for each phantom. This process was repeated for all the images corresponding to different phantom configurations (e.g., different denoising methods).

\section{Experimental Results}
\label{sec 4}
\subsection{Analysis of fMRI Images}
The evaluation of the denoised images using our method was conducted by comparing them to the noise-free samples using two commonly utilized image quality measurements: Peak signal-to-noise ratio (PSNR) \cite{yu2019ea}, and structural similarity index (SSIM) \cite{yu2019ea}.
PSNR serves as a relative estimation of image quality, measured on a logarithmic decibel scale. A higher PSNR value often indicates better synthesis performance. However, PSNR's reliance on the MSE between the noise-free and denoised images makes it susceptible to potential bias caused by excessive smoothing. In contrast, SSIM evaluates the perceived changes in structural information between two images, including edges, and serves as a valuable complement to PSNR. SSIM values range from 0 to 1, with higher values indicating a greater similarity between the two images.
Given a noise-free image $I_{NF}$ and denoised sample $I_G$, PSNR is defined as follows:

\begin{equation}
PSNR(I_{NF},I_G)=10\log_{10}\frac{V_{total}Max^2(I_{NF},I_G)}{\Vert{I_{NF}}-I_G\Vert_2^2}
\label{7}
\end{equation}
 
where $Max^2(I_{NF},I_G)$ means the maximum intensity
range of $I_{NF}$ and $I_G$, and $V_{total}$ denotes the total number of voxels of $I_{NF}$ or  $I_G$. SSIM is calculated by the following equation.

\begin{equation}
SSIM(I_{NF},I_G)=\frac{(2\mu_{I_{NF}}\mu_{I_G}+c_1)(2\sigma_{I_{NF}I_G}+c_2)}{(\mu_{I_{NF}}^2+\mu_{I_G}^2+c_1)(\sigma_{I_{NF}}^2+\sigma_{I_G}^2+c_2)}
\label{8}
\end{equation}

where $\mu_{I_{NF}}$, $\mu_{I_G}$, $\sigma_{I_{NF}}$, and $\sigma_{I_G}$ are the means and variances of
image $I_{NF}$ and $I_G$, and $\sigma_{I_{NF}I_G}$ is the covariance of $I_{NF}$ and $I_G$. 

We compared the performance of the proposed method for two different slice-based, 3D UWGAN$_{xyz}$, and time-based, 3D UWGAN$_{xyt}$, configurations. Table \ref{1} shows the PSNR and SSIM metrics for two configurations on CTNI rsfMRI dataset. Since the Rician noise level ($\delta$) in real preclinical data is about 3\% experimentally, inspired by \cite{ran2019denoising}, we applied the noise standard deviation, $\delta$ from 1\% to 9\%, increasing in increments of 2\%. As the table shows the time-based configuration outperforms the slice-based configuration for different levels of the Rician noise.

\begin{table}[ht]
\caption{PSNR and SSIM (mean $\pm$ standard deviation) on CTNI rsfMRI DB} 
\label{1}     
\begin{tabular}{c c c c c c c} 
\hline
\rule[-1ex]{0pt}{3.5ex} Noise ($\delta$)&1\%&3\%&5\%&7\%&9\%\\
\hline
\rule[-1ex]{0pt}{3.5ex} Rician Noise&39.04 $\pm$ 4.22 &28.97 $\pm$ 3.84&24.78 $\pm$ 3.81&21.64 $\pm$ 3.49&18.83 $\pm$ 3.37\\
\rule[-1ex]{0pt}{3.5ex}&0.831 $\pm$ 0.021&0.598 $\pm$ 0.028&0.491 $\pm$ 0.025&0.428 $\pm$ 0.031&0.341 $\pm$0.021\\
\rule[-1ex]{0pt}{3.5ex} BM4D \cite{maggioni2012nonlocal}&42.12 $\pm$ 4.11&35.24 $\pm$ 3.87&32.71 $\pm$ 3.75&31.43 $\pm$ 3.24&31.15 $\pm$ 3.71\\
\rule[-1ex]{0pt}{3.5ex}&0.956 $\pm$ 0.012&0.917 $\pm$ 0.014&0.874 $\pm$ 0.011&0.843 $\pm$ 0.012&0.837 $\pm$ 0.017\\
\rule[-1ex]{0pt}{3.5ex} RED-WGAN \cite{ran2019denoising}&42.87 $\pm$ 3.52&35.75 $\pm$ 3.72&32.19 $\pm$ 3.31&32.28 $\pm$ 3.11&31.45 $\pm$ 3.14\\
\rule[-1ex]{0pt}{3.5ex}&0.960 $\pm$ 0.015&0.922 $\pm$ 0.024&0.883 $\pm$ 0.031&0.852 $\pm$ 0.017&0.844 $\pm$ 0.025\\
\rule[-1ex]{0pt}{3.5ex} 3D GAN \cite{moghari2021efficient}&43.17 $\pm$ 3.85&36.24 $\pm$ 3.61&33.64 $\pm$ 3.54&33.12 $\pm$ 2.76&32.19 $\pm$ 2.85\\
\rule[-1ex]{0pt}{3.5ex}&0.971 $\pm$ 0.024&0.935 $\pm$ 0.031&0.892 $\pm$ 0.028&0.869 $\pm$ 0.019&0.851 $\pm$ 0.021\\
\rule[-1ex]{0pt}{3.5ex} DU-GAN \cite{huang2021gan}&44.07 $\pm$ 2.86&37.58 $\pm$ 2.75&34.15 $\pm$ 2.65&33.89 $\pm$ 1.88&33.25 $\pm$ 1.23\\
\rule[-1ex]{0pt}{3.5ex}&0.974 $\pm$ 0.023&0.936 $\pm$ 0.020&0.901 $\pm$ 0.019&0.876 $\pm$ 0.022 &0.857 $\pm$ 0.023\\
\rule[-1ex]{0pt}{3.5ex} RIRGAN \cite{yu2023rirgan}&44.53 $\pm$ 2.28&37.70 $\pm$ 2.75&34.23 $\pm$ 2.57&33.90 $\pm$ 1.69&33.31 $\pm$ 1.19\\
\rule[-1ex]{0pt}{3.5ex}&0.975 $\pm$ 0.021&0.938 $\pm$ 0.021&0.903 $\pm$ 0.017&0.878 $\pm$ 0.024 &0.859 $\pm$ 0.021\\
\rule[-1ex]{0pt}{3.5ex} 3D U-WGAN$_{xyz}$&44.81 $\pm$ 2.31&37.93 $\pm$ 2.63&34.98 $\pm$ 2.44&34.02 $\pm$ 1.50&33.87 $\pm$ 1.17\\
\rule[-1ex]{0pt}{3.5ex}(ours)&0.978 $\pm$ 0.015&0.940 $\pm$ 0.017&0.908 $\pm$ 0.012&0.881 $\pm$ 0.012&0.861 $\pm$ 0.019\\
\rule[-1ex]{0pt}{3.5ex} 3D U-WGAN$_{xyt}$&\textbf{45.78 $\pm$ 2.73}&\textbf{38.34 $\pm$ 2.58}&\textbf{36.10 $\pm$ 1.75}&\textbf{35.14 $\pm$ 1.89}&\textbf{35.08 $\pm$ 1.03}\\
\rule[-1ex]{0pt}{3.5ex}(ours)&\textbf{0.981 $\pm$ 0.011}&\textbf{0.953 $\pm$ 0.010}&\textbf{0.912 $\pm$ 0.009}&\textbf{0.908 $\pm$ 0.011}&\textbf{0.895 $\pm$ 0.012}\\
\hline
\end{tabular}
\end{table}

We also compared the performance of the proposed method with block-matching and 4D filtering (BM4D) \cite{maggioni2012nonlocal} as widely used conventional method for MRI denoising. The BM4D algorithm is an extension of the BM3D filter to volumetric data. It operates by grouping mutually similar 3D patches from the image and collectively applying filtering techniques within a 4D array in a transformed domain. Also, we compared our work with GAN-based methods proposed for denoising CT and MR images including RED-WGAN \cite{ran2019denoising}, 3D GAN \cite{moghari2021efficient} which 3D U-net generator with skip connections is its core component, DU-GAN \cite{huang2021gan} developed for denoising 2D CT images with U-net based discriminator, and RIRGAN denoising model \cite{yu2023rirgan} which applied GAN using residual-in-residual-blocks (RIR-Blocks) for denoising low resolution MRI data.  
The results in Table \ref{1} indicate that our denoising method with time-based configuration, 3D UWGAN$_{xyt}$, achieves superior performance compared to the conventional and deep learning based techniques in all evaluation measures.

Table \ref{3} shows the PSNR and SSIM metrics on CTNI task fMRI dataset.
We compared our method with BM4D \cite{maggioni2012nonlocal}, 3D GAN \cite{moghari2021efficient}, DU-GAN \cite{huang2021gan}, and RIRGAN \cite{yu2023rirgan} methods on CTNI task fMRI data. 
Since the time-based configuration outperformed the slice-based configuration in the previous test, in this part we reported results related to time-based model, 3D U-WGAN$_{xyt}$ and compared them with the state-of-the-art methods for CTNI task fMRI data. 

\begin{table}[ht]
\caption{PSNR and SSIM (mean $\pm$ standard deviation) on CTNI task fMRI DB} 
\label{3}      
\begin{tabular}{c c c c c c c}
\hline
\rule[-1ex]{0pt}{3.5ex} Noise ($\delta$)&1\%&3\%&5\%&7\%&9\%\\
\hline
\rule[-1ex]{0pt}{3.5ex} Rician Noise&38.12 $\pm$ 3.86 &28.15 $\pm$ 3.56&23.48 $\pm$ 3.79&21.12 $\pm$ 3.38&17.59 $\pm$ 3.42\\
&0.828 $\pm$ 0.031&0.579 $\pm$ 0.029&0.476 $\pm$ 0.029&0.431 $\pm$ 0.034&0.329 $\pm$0.029\\
\rule[-1ex]{0pt}{3.5ex} BM4D \cite{maggioni2012nonlocal}&40.11 $\pm$ 4.15&34.71 $\pm$ 3.92&31.15 $\pm$ 3.82&30.62 $\pm$ 3.63&29.98 $\pm$ 3.82\\
&0.934 $\pm$ 0.023&0.908 $\pm$ 0.021&0.869 $\pm$ 0.023&0.832 $\pm$ 0.026&0.828 $\pm$ 0.022\\
\rule[-1ex]{0pt}{3.5ex} 3D GAN \cite{moghari2021efficient}&40.09 $\pm$ 3.92&35.11 $\pm$ 3.75&32.18 $\pm$ 3.62&32.35 $\pm$ 2.69&31.08 $\pm$ 2.64\\
&0.952 $\pm$ 0.027&0.916 $\pm$ 0.032&0.873 $\pm$ 0.031&0.844 $\pm$ 0.023&0.836 $\pm$ 0.025\\
\rule[-1ex]{0pt}{3.5ex} DU-GAN \cite{huang2021gan}&41.14 $\pm$ 2.54&36.72 $\pm$ 2.64&33.81 $\pm$ 2.45&32.98 $\pm$ 1.75&31.89 $\pm$ 1.74\\
&0.961 $\pm$ 0.025&0.920 $\pm$ 0.029&0.884 $\pm$ 0.023&0.851 $\pm$ 0.021 &0.842 $\pm$ 0.023\\
\rule[-1ex]{0pt}{3.5ex} RIRGAN \cite{yu2023rirgan}&42.24 $\pm$ 2.31&36.94 $\pm$ 2.49&34.23 $\pm$ 2.57&33.26 $\pm$ 1.58&32.40 $\pm$ 1.24\\
&0.969 $\pm$ 0.022&0.923 $\pm$ 0.027&0.893 $\pm$ 0.019&0.864 $\pm$ 0.020 &0.848 $\pm$ 0.018\\
\rule[-1ex]{0pt}{3.5ex} 3D U-WGAN$_{xyt}$&\textbf{43.19 $\pm$ 2.27}&\textbf{37.17 $\pm$ 2.32}&\textbf{35.41 $\pm$ 1.96}&\textbf{34.37 $\pm$ 1.44}&\textbf{33.19 $\pm$ 1.09}\\
(ours)&\textbf{0.973 $\pm$ 0.021}&\textbf{0.953 $\pm$ 0.010}&\textbf{0.902 $\pm$ 0.016}&\textbf{0.878 $\pm$ 0.019}&\textbf{0.856 $\pm$ 0.011}\\
\hline    
\end{tabular}
\end{table}

\begin{figure}
\centering
\begin{tabular}{c}
\includegraphics[height=8.5cm]{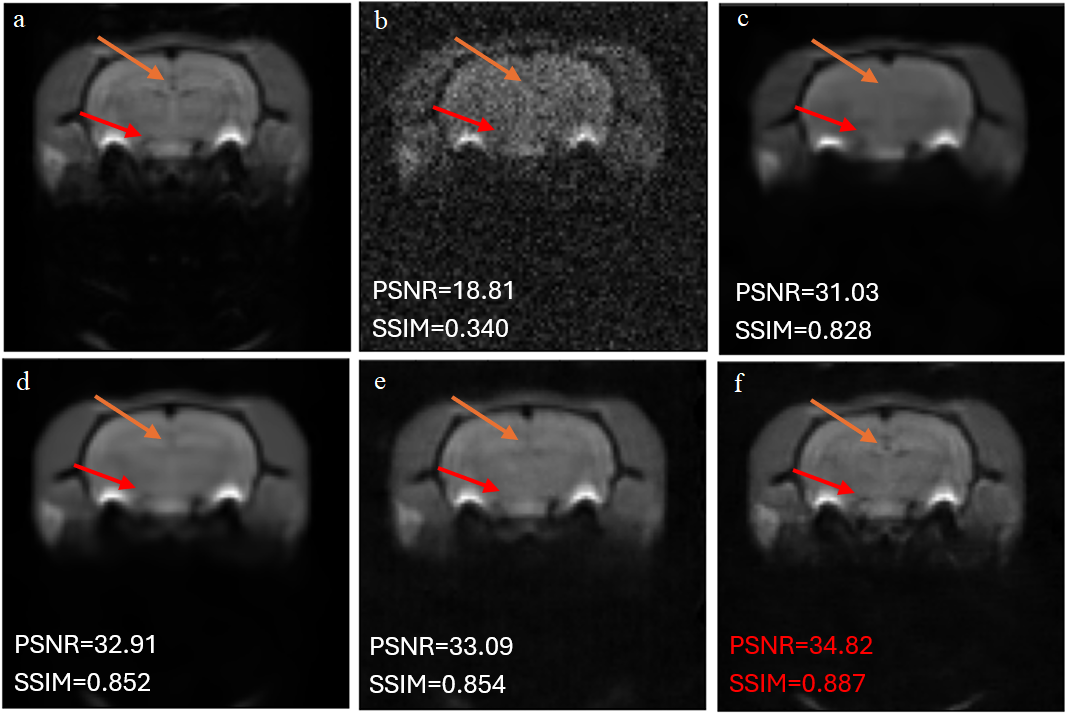}
\end{tabular}
\caption 
{ \label{Fig. 5}
Denoised example from CTNI rsfMRI DB (a) Noise-free image, (b) Noisy image (noise level: 9\%), (c) BM4D \cite{maggioni2012nonlocal}, (d) DU-GAN \cite{huang2021gan}, (e) RIRGAN \cite{yu2023rirgan}, (f) 3D U-WGAN$_{xyt}$ (ours).}
\end{figure}

\begin{figure}
\begin{center}
\begin{tabular}{c}
\includegraphics[height=8.5cm]{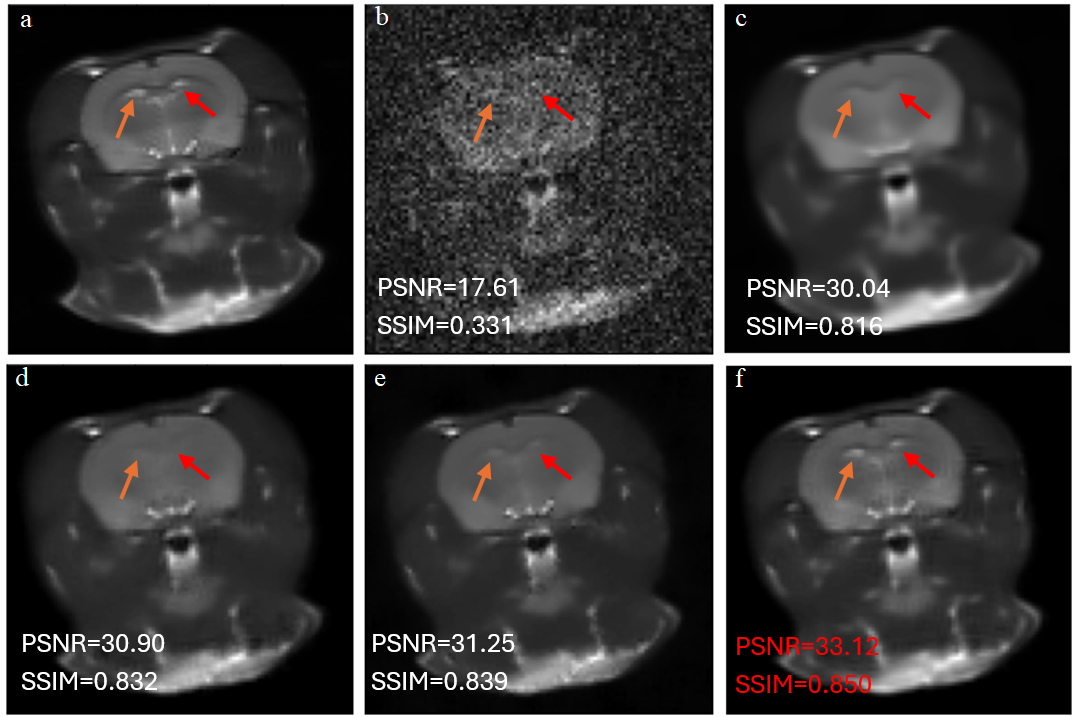}
\end{tabular}
\end{center}
\caption 
{ \label{Fig. 6}
Denoised example from CTNI task fMRI DB (a) Noise-free image, (b) Noisy image (noise level: 9\%), (c) BM4D \cite{maggioni2012nonlocal}, (d) DU-GAN \cite{huang2021gan}, (e) RIRGAN \cite{yu2023rirgan}, (f) 3D U-WGAN$_{xyt}$ (ours).} 
\end{figure}

The qualitative results of applying the proposed denoising method on the samples from CTNI rsfMRI, and task fMRI datasets and comparison with other methods have been demonstrated in Fig. \ref{Fig. 5}, and Fig. \ref{Fig. 6} respectively. The testing samples were corrupted with 9\% Rician noise to demonstrate the effectiveness of our proposed denoising method.
We showed that the time-based configuration outperforms the slice-based, all results related to the proposed method has been done under time-based configuration.
From the regions indicated by orange and red arrows, it is
obvious that the 3D U-WGAN$_{xyt}$ method performs better than the BM4D, DU-GAN, and RIRGAN methods in terms of structure preservation for CTNI rsfMRI and task fMRI datasets. 
All of the methods employed were able to reduce the noise to varying extents. However, BM4D exhibited over smoothing effects and distorted important details, as evident in these figures. The DU-GAN and RIRGAN algorithms also suffer from some information loss. 

Results show that the 3D U-WGAN$_{xyt}$ method achieved highest performance by avoiding over smoothing and retaining a higher level of structural details.
This advantage is derived from the 3D U-Net-based discriminator in our proposed algorithm and spatial-temporal configuration of 4D data. This discriminator offers feedback on both global structures and local details to the generator. 
Moreover, this advantage is due to the incorporation of WGAN, dense U-Net, and the combined loss functions, which effectively generates results that closely resemble the original data distribution.

Based on the above discussion, in terms of noise suppression, 3D U-WGAN$_{xyt}$ outperformed other methods consistently produced results that were closer to the reference images. Quantitative results obtained from various methods have been shown on each figure. The findings indicate that 3D U-WGAN$_{xyt}$ exhibits the highest performance in terms of PSNR, and SSIM, aligning with the observations made through visual inspection.
Results in this section show that our proposed method can be applied on both rsfMRI and task fMRI data. 

\subsection{Evaluation of Denoising Techniques in Simulated fMRI Data}
In this part, we evaluate brain functional activity before and after applying denoising techniques to compare our proposed algorithm with state-of the-art methods. The preprocessing pipeline explained in section ~\ref{sec 3.3} was repeated for three rat subjects under the following conditions, (1) without denoising, (2) with BM4D denoising, (3) with RIRGAN denoising, (4) using DU-GAN denoising, and (5) using 3D U-WGAN denoising.

We focused on comparing the performance of various denoising techniques across several regions of interest (ROIs) in the rat brain during fMRI phantom generation. Specifically, we analyse the percentage of statistically significant voxels detected in each ROI, comparing the performance across the denoising methods.

Our aim was to evaluate how well each method restored activation patterns in key brain regions relative to the ground truth. As mentioned, the key ROIs, namely the superior colliculus and both visual cortices, represent critical areas of the brain involved in visual processing in rodents. These regions were chosen due to their well-defined activation patterns in both real and simulated fMRI studies.

By quantifying the percentage of statistically significant voxels in each ROI, we could directly assess the effect of denoising on the recovery of activation patterns. The ground truth percentages represent the ideal distribution of activations, and deviations from these values across the different denoising methods provide insights into the trade-offs between noise removal and preservation of true activation signals.

We employed this region-based analysis to provide a comprehensive assessment of each denoising method's effectiveness in restoring fMRI activation patterns across multiple ROIs. This approach allows for a robust evaluation of the denoising techniques by comparing their performance not only in terms of overall signal recovery but also in their ability to preserve activations in specific brain regions with varying signal intensities and noise levels. By focusing on distinct ROIs, we can systematically assess the spatial fidelity of each denoising method and its potential impact on the accuracy of functional connectivity analyses and brain network characterization.

\begin{figure}
\begin{center}
\begin{tabular}{c}
\includegraphics[height=10cm]{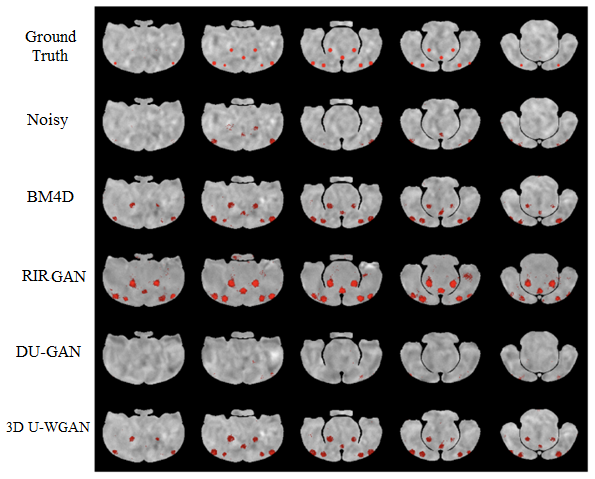}
\end{tabular}
\end{center}
\caption 
{ \label{Fig. 7} Simulating phantom fMRI activation patterns after applying various denoising methods.
}
\end{figure}
Figure \ref{Fig. 7} demonstrates the superior performance of 3D-UWGAN in accurately detecting activated voxels. The top row represents the ground truth, with red regions indicating known activation areas. In the second row, without denoising, noise significantly obscures the true activation patterns. The BM4D method (third row) enhances sensitivity, detecting a broader range of activations but perhaps at the cost of numerous false positives. RIRGAN method (fourth row) further increases sensitivity but introduces excessive false positives, diverging significantly from the ground truth. Interestingly, DU-GAN (fifth row) detects even fewer activated voxels than the noisy case, potentially missing critical activations due to over-filtering. In contrast, 3D-UWGAN (bottom row) demonstrates superior performance, closely matching the ground truth while minimizing false positives. This result highlights 3D U-WGAN's ability to maintain a precise balance between sensitivity and specificity, outperforming both traditional methods and other deep learning approaches by effectively preserving true activations and filtering out noise.

\begin{figure}
\begin{center}
\begin{tabular}{c}
\includegraphics[height=6cm]{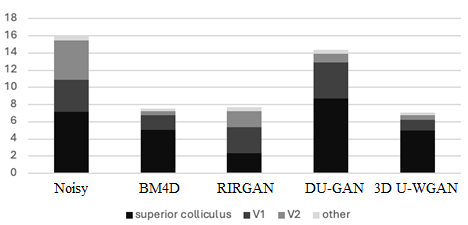}
\end{tabular}
\end{center}
\caption 
{ \label{Fig. 8} Weighted deviation of significant voxel detection from ground truth activation across various denoising methods.
}
\end{figure}

Weighted deviations of activation from ground truth across various denoising methods is shown in Fig. \ref{Fig. 8}. The x-axis represents the different denoising methods applied to the simulated fMRI data, and the y-axis shows the weighted average deviation of statistically significant voxels from the ground truth activation patterns. Deviations are weighted by the percentage of ground truth voxels each region of interest (ROI) occupies. Lower values indicate a closer match between the detected activation and the ground truth, with each method's performance visually represented by the height of the stacked bars.

\begin{figure}
\begin{center}
\begin{tabular}{c}
\includegraphics[height=10cm]{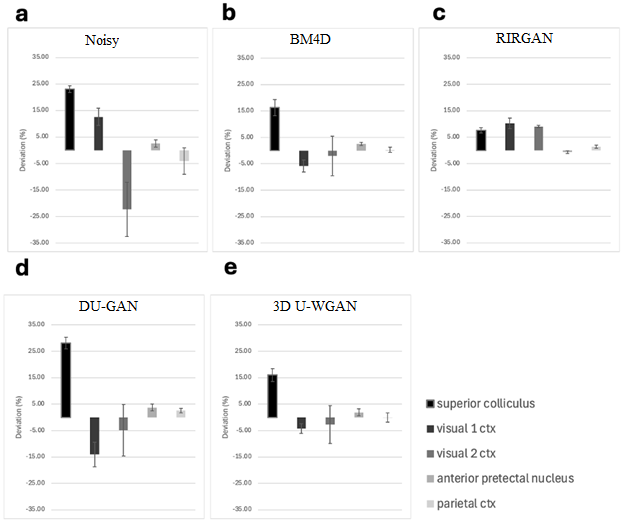}
\end{tabular}
\end{center}
\caption 
{ \label{Fig. 9} Percentage deviation from ground truth in significant voxel detection across five denoising methods for different regions of interest (ROIs).
} 
\end{figure}

Figure \ref{Fig. 9} illustrates percentage deviation from ground truth in significant voxel detection across five denoising methods for different regions of interest (ROIs). Each panel represents a different denoising method, with deviations plotted for the superior colliculus, visual cortex 1 (V1), visual cortex 2 (V2), anterior pretectal nucleus, and parietal cortex. The y-axis indicates the percentage deviation from ground truth, with positive values representing overestimation and negative values representing underestimation of detected voxels. Error bars represent the variability across multiple simulations.
Panel a (Noisy): Non-denoised data shows significant overestimation in the superior colliculus and underestimation in V2.
Panel b (BM4D): BM4D reduces deviations across the three main ROIs (superior colliculus, V1, and V2) compared to the non-denoised case. There remained a high standard error with the estimation of V2. 
Panel c (RIRGAN): RIRGAN shows minimal deviations across all ROIs, with slight overestimation in superior colliculus and both visual cortices, and minimal underestimation in other regions.
Panel d (DU-GAN): DU-GAN demonstrates a significant overestimation in the superior colliculus and substantial underestimation in visual cortex 1 and visual cortex 2.
Panel e (3D U-WGAN): 3D U-WGAN improves deviation in visual cortex 2 and parietal cortex, though there remains overestimation in superior colliculus and slight underestimation in other regions.
 
\subsection{Seed-based Functional Connectivity Analysis}
\begin{figure}
\begin{center}
\begin{tabular}{c}
\includegraphics[height=15cm]{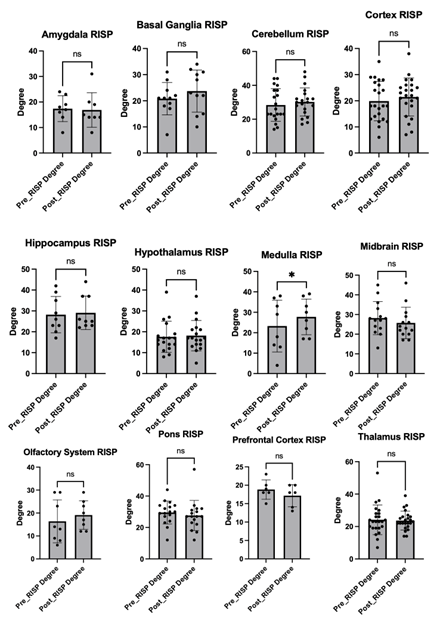}
\end{tabular}
\end{center}
\caption 
{ \label{Fig. 10} Seed-based Functional Connectivity on CTNI noisy data.
} 
\end{figure}

\begin{figure}
\begin{center}
\begin{tabular}{c}
\includegraphics[height=12cm]{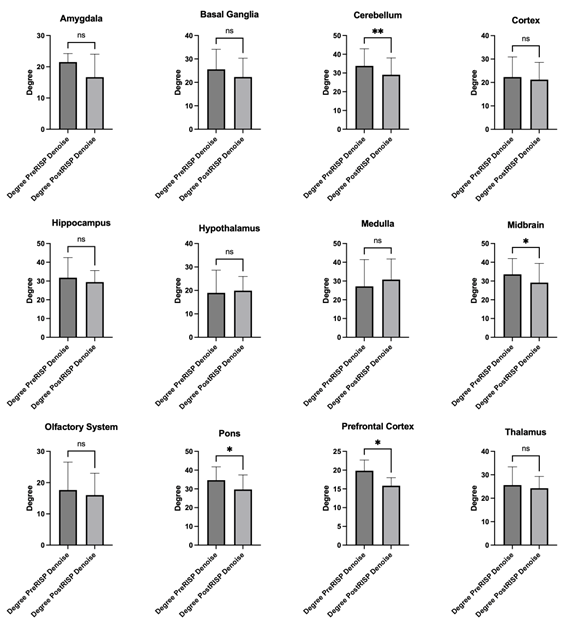}
\end{tabular}
\end{center}
\caption 
{ \label{Fig. 11} Seed-based Functional Connectivity on CTNI denoised data using proposed method.
} 
\end{figure}

We performed seed-based functional connectivity to illustrate denoising effects on downstream analysis on real CTNI dataset. We compared the results with non-denoised data and with denoised data on a neuroimaging study whose goals were to assess changes in brain activity associated with aggression as well as the effectiveness of drug intervention in awake rats.

Aggression was tested in male rats by introducing an "intruder" rat, a novel male rat, near the male rat. Aggression was verified by observing the onset of piloerection of the fur along the lower midline back. Functional scans were taken prior to introducing the "intruder" rat, as well as post.

Risperidone (RISP) was one of the drugs profiled. The denoising identified new regions that differed between no-intruder and intruder conditions while simultaneously eliminating one region from being significantly different. Specifically, the denoised data showed significant differences between pre and post risperidone in the midbrain, pons and prefrontal cortex at p$<$0.05 level, and cerebellum at the p$<$0.01 level. The medulla was significantly different in the non-denoised data, but not so in the denoised data. These results are shown in Fig. \ref{Fig. 10} and Fig. \ref{Fig. 11} for noisy and denoised data respectively. These preliminary results indicate that denoising may help in discerning differences in group level analysis in functional connectivity.

\subsection{Ablation Study}
In this section, we discuss the ablation study we performed to thoroughly investigate our proposed method's efficacy concerning various aspects, including the significance of individual components, dense U-Net based discriminator, loss functions, and different patch sizes. The ablation study was executed using the testing set of the CTNI rsfMRI dataset with Rician noise level of 9\%.

\begin{table}[ht]
\caption{Ablation study on CTNI rsfMRI test set. (PSNR and SSIM (mean $\pm$ standard deviation))} 
\label{7}    
\begin{tabular}{c c c c c c c c}
\hline
\rule[-1ex]{0pt}{3.5ex} Method&PSNR &SSIM \\
\hline
\rule[-1ex]{0pt}{3.5ex} Baseline (WGAN+$L_{MSE}$)&31.11$\pm$1.79&0.841$\pm$0.007\\
\rule[-1ex]{0pt}{3.5ex} Baseline+$L_{Per}$&32.15$\pm$2.24&0.848$\pm$0.017\\
\rule[-1ex]{0pt}{3.5ex} Baseline+$L_{Per}$+3D U-Net discriminator+$L_{D}$&34.52$\pm$1.18&0.881$\pm$0.015\\
\rule[-1ex]{0pt}{3.5ex} Baseline+$L_{Per}$+3D dense U-Net discriminator+$L_{D}$&\textbf{35.08$\pm$1.03}&\textbf{0.895$\pm$0.012}\\
\hline
\end{tabular}
\end{table}

Table \ref{7} compares quantitative metrics of combination of different components with baseline method. The baseline method is WGAN which includes the traditional classification discriminator. In the baseline method we applied time-based configuration to handle 4D fMRI data based on the results from previous sections. Initially, substituting the traditional classification discriminator with a 3D U-Net-based discriminator offers the generator both global structural information and local per-pixel feedback concurrently. This results in a substantial improvement in SSIM. To further improve denoising performance, a 3D dense U-Net-based discriminator has been applied.
Also, we set up three different loss functions to evaluate the performance of the proposed method. As the Table \ref{7} shows adding perceptual and discriminator losses to MSE loss could improve system's performance in both PSNR and SSIM metrics. Moreover, because of the U-Net structure of the discriminator, it is crucial to examine how the patch size during training impacts the overall outcome. Since the size of fMRI samples in the CTNI dataset is $96\times96\times22\times300$. We compared two patch sizes including $32\times32$, and $64\times64$.
In this test, we compared both time-based and slice-based data configuration. The results presented in Table \ref{8} indicate that superior performance can be achieved with a smaller patch size and time-based configuration. Larger patch sizes, on the other hand, may pose challenges during training due to a scarcity of training samples. 

\begin{table}[ht]
\caption{Ablation study of patch sizes on CTNI rsfMRI test set. (PSNR and SSIM (mean $\pm$ standard deviation))} 
\label{8}       
\begin{tabular}{c c c c c c c}
\hline
\rule[-1ex]{0pt}{3.5ex} Proposed model&Patch size&PSNR &SSIM\\
\hline
\rule[-1ex]{0pt}{3.5ex} 3D U-WGAN$_{xyt}$&$32\times32$&\textbf{35.08$\pm$1.03}&\textbf{0.895$\pm$0.012}\\
\rule[-1ex]{0pt}{3.5ex} &$64\times64$&34.46$\pm$1.18&0.862$\pm$0.017\\
\rule[-1ex]{0pt}{3.5ex} 3D U-WGAN$_{xyz}$&$32\times32$&33.87$\pm$1.17&0.861$\pm$0.019\\
\rule[-1ex]{0pt}{3.5ex} &$64\times64$&32.21$\pm$2.21&0.853$\pm$0.021\\
\hline
\end{tabular}
\end{table}

\subsection{Computational Cost}
The noteworthy aspect to consider is the computational expense associated with the deep learning-based approach. The most resource-intensive phase is the training process. Although training typically takes place on a GPU to expedite the process, it remains a time-consuming endeavour. In our case, when we alternate between training the generator and discriminator networks for each epoch in our training set, it consumes roughly 20 minutes per epoch for the CTNI database. Conversely, other techniques like BM4D do not require training, but their execution times are considerably longer when compared to deep learning-based methods.

In our study, the average execution times for each subject of the CTNI rsfMRI dataset were 54.74 seconds for BM4D, 45.37 seconds for 3D GAN, 
33.10 seconds for RIRGAN method and 24.18 seconds for 3D-UWGAN$_{xyt}$. It's important to note that, in practice, the runtime of deep learning-based methods can be further reduced by leveraging GPUs for testing.

\section{Conclusion and Future Work}
\label{sec 5}
In this paper, we proposed a new approach called 3D U-WGAN for denoising preclinical fMRI scans. Our method includes a U-Net-based discriminator that provides per-pixel feedback to the denoising network while also focusing on the overall structure of the image. Additionally, we incorporated adversarial loss along with perceptual and MSE losses to improve the denoising algorithm's performance. Through extensive experiments involving visual and quantitative comparisons, we demonstrated the effectiveness of our proposed method.
The paper's results are promising and effectively showcase how GAN can be used to denoise preclinical resting state and task fMRI scans. 
In future work, we plan to make our method more general by extending it to handle different artifacts related to motions.

\bibliography{sn-bibliography}

\end{document}